\documentclass[lettersize,journal]{IEEEtran}
\usepackage{amsmath,amsfonts}
\usepackage{algorithmic}
\usepackage{algorithm}
\usepackage{array}
\usepackage[caption=false,font=normalsize,labelfont=sf,textfont=sf]{subfig}
\usepackage{textcomp}
\usepackage{stfloats}
\usepackage{url}
\usepackage{verbatim}
\usepackage{graphicx}
\usepackage{cite}
\def\BibTeX{{\rm B\kern-.05em{\sc i\kern-.025em b}\kern-.08em
    T\kern-.1667em\lower.7ex\hbox{E}\kern-.125emX}}
\usepackage{balance}
\usepackage{placeins}
\usepackage{float}
\usepackage{svg}
\usepackage{authblk}
\usepackage{epsfig}
\usepackage{amsmath}
\usepackage{float}
\usepackage{booktabs} 
\begin{document}
\title
{\color{black}
Maximal electric power generation from varying 
ocean waves with LC-tuned reactive PTO force}
\author{Jingxin Zhang*, Uzair Bin Tahir*, Richard Manasseh \vspace{-2em}

\thanks{ This work was funded by ACSRF-WWE and performed at Swinburne University of Technology, Australia.\\
J. Zhang and U. Tahir (*equal first authors) are with the School of Science, Computing and Engineering Technologies, Swinburne University of Technology, 3122, Victoria, Australia (jingxinzhang, utahir@swin.edu.au).\\
R. Manasseh is with the School of Engineering, Swinburne University of Technology, 3122, Victoria, Australia (rmanasseh@swin.edu.au).}
}
\markboth{IEEE Transactions on Sustainable Energy, March~2024}%
{Shell \MakeLowercase{\textit{et al.}}: A Sample Article Using IEEEtran.cls for IEEE Journals}
\maketitle
\begin{abstract}
The reactive Power Take Off (PTO) force is the key to maximizing  mechanical power absorption and electric power generation of Wave Energy Converters (WECs) from ocean waves with variable frequency, but its study is limited due to its difficulty in physical realization. This paper presents a simple yet effective $LC$-tuned WEC that generates a tunable reactive PTO force from tunable inductor $L$ and capacitor $C$ in the WEC. A complete closed loop system model of the WEC is derived first, then three quantitative rules are obtained from analyzing the model. These rules are used to tune the $LC$ network, and hence the reactive PTO force that drives the WEC, to resonate with the input wave force and generate maximal electric power over a range of wave frequencies. Mathematical analysis of the WEC and tuning rules reveals the analytical and quantitative descriptions of the WEC's mechanical power absorption, active and reactive electric power generation and power factor, optimal electric resistance load, and the generator and $LC$ capacity requirements. Simulation results show the effectiveness and advantages of the proposed WEC and verify the analysis results. 
\end{abstract}
\vspace{-1em}
\section{Introduction}
{\color{black} 
\IEEEPARstart{W}{ave} energy converters (WECs) play the central role in ocean wave power generation, converting wave energy to electric energy. Among various types of WECs, the point absorber WEC 
is one of the simplest and most promising types \cite{manasseh2021fluid} and is the WEC addressed in this paper.

A WEC absorbs more power to produce electricity when it resonates with incoming waves. But this is extremely difficult to sustain under varying waves and electric loads \cite{ringwood2014energy, manasseh2021fluid, korde2016hydrodynamic}. 
To solve this problem, the power take off (PTO) system of the WEC is used to maximize power absorption in such situations. 

The PTO system is a crucial part of a WEC. 
It absorbs mechanical power from the WEC's oscillator, generates the PTO force against the wave input force to sustain oscillation and mechanical power absorption, and converts absorbed mechanical power to electric power with the WEC's generator. Its design and operation directly affect the mechanical power absorption and electric power generation of the WEC \cite{wang2018review}. 

In attempts to make a WEC resonate over a range of frequencies around its natural frequency, many mechanical PTO methods have been proposed in the literature that can `tune' the WEC, 
e.g., 
hydraulic PTOs \cite{forehand2015fully}, reactive actuators 
\cite{
genest2014effect}, auxiliary mass control \cite{khasawneh2021internally}, impedance matching \cite{sugiura2020wave}, latching control \cite{henriques2016latching}, 
geometry and position control\cite{babarit2010assessment} and conventional spring-damper systems \cite{gu2021power}. These methods do not address electric power generation directly, and generally require additional mechanical and electrical devices and external power injection, 
 on top of the generator and wave input power, to generate and control the PTO force. 
Hence, they tend to be
complex,  expensive, inefficient and less responsive,   and may lower the net (electric) power output of the WEC.

In attempts to directly maximize the WEC's electric power generation under varying waves,  many electrical PTO methods have also been proposed, 
e.g.,  
electric resistive latching \cite{tahir2023latching}, impedance tuning \cite{guo2016continuous},
 damping control \cite{park2016active}, reactive control \cite{tedeschi2011effect}, and maximum power point
tracking \cite{hardy2016maximum,xu2021maximum}. 
These methods use the WEC's generator to generate and control the PTO force while producing electricity,
without additional devices and external power injection. Thus, they are simpler, lower cost, more  responsive and  efficient, and may 
maximize the net (electric) power output of the WEC. 

Despite their advantages, the electrical methods share a common problem: they only focus on the generation of active electric power and overlook the reactive electric power required by such generation. Many of them use the d-q axis generator model and intentionally zero or suppress one axis current to block or minimize the generator's reactive current,
e.g., \cite{sergiienko2022comparison,tahir2023latching,
park2016active}. As a result, they generally do not produce reactive PTO forces. As shown in some previous works \cite{korde2016hydrodynamic,
genest2014effect,khasawneh2021internally,sugiura2020wave, jain2022limiting} and also in the sequel, the reactive PTO forces are the key to making the WEC resonate and hence to maximizing electric power generation over a range of wave frequencies. The lack of reactive PTO forces  makes it difficult for the electrical methods to maximize  electric power generation and may result in the negative electric 
power generation observed in \cite{sergiienko2022comparison} and references therein. 
Overlooking the generator's reactive current has also led to the absence of device sizing analyses, e.g. an analysis of generator capacity rating, in the existing works.  

To solve these problems, we propose a new WEC structure with a novel electric PTO system comprising a  permanent magnet linear generator (PMLG) in parallel connection with tuneable inductor $L$ and capacitor $C$, as shown in Fig. 1. The $L$ and $C$ are used to produce respectively the lagging and leading reactive currents in the generator to induce the lagging and leading reactive PTO forces vs the wave input force. By tuning the $L$ and $C$ and hence 
the reactive PTO forces according to the wave frequency, we keep the WEC resonating and its active electric power output constant and maximal over a range of wave frequencies, under ideal conditions. 

To reveal the underlying physics of the proposed WEC, we use the ideal models of 1D ocean waves and linear WECs, and a simplified single phase linear generator with negligible internal resistance and inductance in our derivation and analysis. Also, we only consider steady state operation of the WEC without dynamic  control of the PTO. The proposed 
setup allows the derivation of a complete closed loop system model of the proposed WEC, quantitative and analytical descriptions and conditions for maximum mechanical absorption, maximum active electric power generation,   the active,  reactive and apparent powers and power factor of the WEC's generator, and generator capacity rating in term of apparent power. All of these are previously unknown results to the best of our knowledge. 

The rest of paper is organized as follows. Section II describes the proposed new WEC, its PTO force, and its closed loop system model. Section III presents the $LC$ tuning rules of the new WEC, and
the quantitative analysis and analytical results of these rules. Section IV presents a simulation verification and demonstration of the analytical results. Section V draws the conclusions and discusses the implications, limitations and extensions of the presented results. 
\vspace{-0.2cm}
{\color{black}
\section{Proposed new WEC structure and model}
\begin{figure}[!]
\centering
\includegraphics[width= 8 cm,height=6cm]{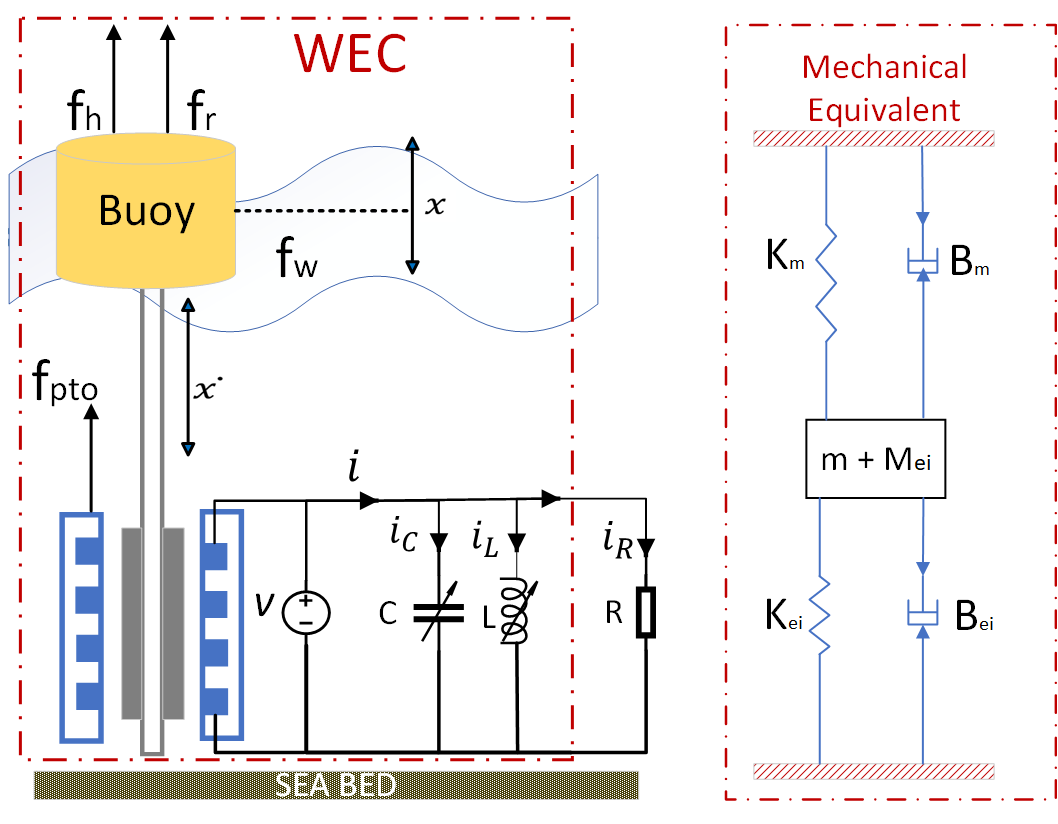}
\caption{{\color{black}New WEC structure and mechanical-electrical interaction model 
}}
\vspace{-1em}
\label{Complete design}
\end{figure}

\subsection{Proposed new WEC structure} 

Fig. \ref{Complete design} (left) shows our proposed new WEC structure.  It has a heaving buoy driven by the ocean waves and connected to the PMLG by a shaft. The PMLG is connected to a tunable parallel $LC$ network that provides electricity to the load (resistor) $R$. This structure is similar to the class of WECs using direct electric drive as the PTO system \cite{samad2022ocean}, but its tunable $LC$ network and PMLG form a new type of PTO system that can be tuned to maximize the electric power output to the load $R$ over a range of wave frequencies. Though simple, the basic principle derived from the proposed WEC is applicable to other classes of WECs \cite{manasseh2021fluid, penalba2016review, ringwood2014energy}, as long as their mechanical 
operation can be modelled as a resonating linear mass-damper-spring system. 

Fig. \ref{Complete design} (right) shows the Mechanical-Electrical (M-E) interactions in the proposed WEC.  
As the buoy oscillates, it moves the PMLG's translator to generate 
the voltage $v(t)$ that drives the current $i(t)$ through the $RLC$ network; the current $i(t)= i_R(t)+i_L(t)+i_C(t)$, in turn, induces the electrical resistive force $f_{pto}(t) = f_R(t)+ f_L(t)+f_C(t)$ with three sub-forces on the shaft of the PMLG; 
the three sub-forces 
exerted through the shaft to the buoy 
induce the equivalent mass $M_{ei}$, damping $B_{ei}$ and stiffness $K_{ei}$ that affect the overall system dynamics.
It is assumed that the input wave frequency may vary;   
 typically, ocean swell waves with periods ${\cal O}(1-10~\rm{s})$, which carry the most energy,  vary over timescales of days \cite{bergsma2022wave}. However, the wave frequency is assumed to be known, and 
the $LC$ network is tuned, in accordance with the wave frequency, to control the current $i(t)$ and hence the $f_{pto}(t)$ to 
achieve resonance at different wave frequencies.

\vspace{-0.5em}
\subsection{Open loop dynamic model of proposed WEC}
}

To reveal the fundamental M-E dynamics, we assume that 
the whole WEC system is linear, the water is homogeneous, inviscid, irrotational and incompressible and the waves are sinusoidal, so the waves are describable by potential-flow theory \cite{manasseh2021fluid, korde2016hydrodynamic}; 
the added mass corresponding to the radiation force is constant \cite{said2022wave, xu2021maximum}; the shaft is rigid and lossless, and hence the buoy and translator (BT) and its added fluid mass move together as a single mass 
and all the masses are lumped into the system
total mass $M_m$. These assumptions encapsulate the essential traits of WECs and are commonly employed in the analysis and design of WECs \cite{A501_7,li2012synthesis}. 


Denote $x(t)$, $\dot x(t)$ and $\ddot x(t)$ the displacement, velocity and acceleration of the BT, respectively. Then, by the linear potential theory and 
Newton's second law of motion, the dynamics of the proposed WEC can be described, as commonly formulated \cite{li2012synthesis}, \cite{falcao2010wave} by

 \begin{equation}
    M_m \ddot x(t) = f_w(t) - f_{pto}(t) - f_r(t) - f_h(t),
    \label{dyneq}
\end{equation}
where $f_w(t)$ is the wave excitation force, $f_r(t)$ the radiation force, $f_h(t)$ the hydrostatic stiffness force, and $f_{pto}(t)$ the resistive force from the PTO system of proposed WEC.

As often assumed in fundamental linear analyses of WECs \cite{A501_7}, 
the excitation force $f_w(t)$ is from regular incident waves, with angular wave frequency $\omega$, constant amplitude $A_w$ and no phase difference, and takes the form
\begin{equation}
    f_w(t)= A_w \cos(\omega t).
    \label{Fe}
\end{equation}
It is well known that $A_w$ is proportional to the total mass $M_m$, the larger the $M_m$, the higher the $A_w$ \cite{manasseh2021fluid}.
The radiation force $f_r(t)$ (the fluid force due to the oscillating body) is approximated by Cummins' equation \cite{cummins1962impulse} as 
\begin{equation}
\label{Frad}
    f_r(t)  =  B_m \dot x(t),
\end{equation}
where  $B_{\text m}$ is the radiation damping constant due to the interaction of the buoy motion and surrounding water. 

The hydrostatic stiffness force $f_h(t)$ 
is described as \cite{said2022wave}, \cite{rahoor2020comparison}
\begin{equation}
    f_h(t)= K_m x(t), 
    \label{fhyst}
\end{equation}
where $K_m=\rho g S$ is the hydrostatic stiffness, with $\rho$ the water density, $g$ the acceleration due to gravity and $S$ the cross-sectional area of buoy. 

With the $f_r(t)$ and $f_h(t)$ given above, 
the hydrodynamic equation \eqref{dyneq} of the WEC can be written as 
\begin{equation}
    M_m \ddot x(t)+B_m \dot x(t)+ K_m x(t) = f_w(t) - f_{pto}(t),
    \label{hydrodynamic_forces}
\end{equation}
where $f_r(t) = B_m \dot x(t)$ and $f_h(t) = K_m x(t)$ are moved to the left hand side since they are the feedback forces induced by $\dot x(t)$ and $x(t)$, 
while 
$f_w(t)$ and $f_{pto}(t)$ are kept on the right since $f_w(t)$ is an independent external force from waves, and the nature of $f_{pto}(t)$ is unknown and to be analyzed. 

\vspace{-1em}
\subsection{Analysis of 
$f_{pto}(t)$ and closed loop dynamic model of proposed WEC}
\label{Sectpto}

To analyze the $f_{pto}(t)$ force of the proposed WEC system, we assume the PMLG in Fig \ref{Complete design} (left) is a 1-phase machine with $N_p = 2$ poles that generates a sinusoidal output voltage when its translator is in steady reciprocal motion. Examples of such PMLGs can be found in  \cite{song2019design} and the references therein. Since the internal resistance $R_{in}$ and inductance $L_{in}$ of a generator are generally much smaller than the external load $R$ and $L$ \cite{ryff1994electric}, we neglect $R_{in}$ and $L_{in}$ to simplify discussion without loss of essence. 
Hence, the PMLG's internal voltage $v(t)$ equals its terminal output voltage and the oscillation of $v(t)$ follows that of the translator, with the same frequency. 

By electrical machine theory \cite{ryff1994electric}, when the PMLG's translator moves at the velocity $\dot x(t)$, driven by the input force $f_w(t)$, an internal voltage $v(t)$ is induced in its stator winding, with  
\begin{equation}
    v(t) = K_e \dot x(t),
    \label{e}
\end{equation}
where $K_e$ is the electric constant. The internal voltage $v(t)$ then drives a current $i(t)$ through the PMLG and the $RLC$ network. 
The current 
$i(t)=i_R(t)+i_L(t)+i_C(t)$, in turn, induces a resistive force comprising three components on the translator shaft,
\begin{eqnarray}
    f_{ei}(t) = K_t i(t) = K_t i_R(t)+K_t i_L(t)+K_ti_C(t) \label{f} 
\end{eqnarray}
where $K_t$ is the force constant satisfying $K_t = K_e$, although their physical units are different \cite{ryff1994electric}.

Using \eqref{e}, 
$i_R(t) = v(t)/R$, $i_L(t) = \frac{1}{L} \int v(t) dt$ and  $i_C(t) = C dv(t)/dt$,
the $RLC$ currents  
can be written as 
\begin{equation}
i_R(t) = \frac{{K_e \dot x(t)}}{R}, \,\,
i_L(t) =  \frac{K_e}{L}x(t), \,\,
i_C(t) = C K_e  \ddot x(t).
    \label{RLC}
\end{equation}
Define the electrically induced mass, damping and stiffness
\begin{equation}
\label{MBK}
    M_{ei} := K_tK_e C, \,\, B_{ei} := K_tK_e/R, \,\, K_{ei} :=  K_tK_e/L.
\end{equation} 
Then by \eqref{e}-\eqref{MBK}, 
the electrically induced resistive force $f_{ei}(t)$ and its 
three sub-forces can be written as
\begin{eqnarray}
    f_{ei}(t) &= & K_t i(t)  
    = M_{ei} \ddot x(t) + B_{ei} \dot x(t) + K_{ei} x(t)\label{pto force}
\\ 
f_C(t)&:=&K_ti_C(t) = M_{ei} \ddot x(t), \mbox{  (reactive leading)}  \label{fc}\\
f_R(t) &:=& K_t i_R(t) =B_{ei} \dot x(t),\mbox{  (active)} \label{fr}\\ f_L(t)&:=&K_ti_L(t)=K_{ei} x(t).\mbox{ (reactive lagging)}
\label{fl}    
\end{eqnarray}
As shown in Section \ref{Sect 3}, $f_R(t)$ is the active sub-force; $f_C(t)$ and $f_L(t)$ are respectively the reactive leading and lagging  sub-forces absorbing zero average mechanical power.

Since $f_{ei}(t)$ is the only force from the PMLG against the input force $f_w(t)$, 
$f_{ei}(t)=f_{pto}(t)$ in \eqref{hydrodynamic_forces}. Substituting \eqref{pto force} into \eqref{hydrodynamic_forces} 
gives the closed-loop dynamic 
model of the proposed WEC
\begin{eqnarray}
\label{hydrodynamic_A}
 (M_m+M_{ei})\ddot x(t) + (B_m+B_{ei})\dot x(t) + \hspace{1cm} \nonumber\\ 
 (K_{m}+ K_{ei})x(t) = f_w(t). 
\end{eqnarray}
As shown below,   \eqref{hydrodynamic_A} is a passive linear system with bounded input and bounded output (BIBO) stability and hence admits the Fourier transform (on both sides) \cite{kwakernaak1991modern}
\begin{eqnarray}
\label{FT}
\left[(j\omega)^2 (M_m+M_{ei}) + j\omega(B_m+B_{ei}) +  \right. \hspace{1cm} \nonumber\\ 
\left.(K_m+ K_{ei})\right] X(j\omega) = F_w(j\omega).
\end{eqnarray}
Rearranging \eqref{FT} and defining the frequency response function 
\begin{equation} 
\label{FTF}
H(j\omega) := \frac{1}{(K_m+ K_{ei})-\omega^2(M_m+M_{ei}) + j\omega (B_m+B_{ei})}
\end{equation} 
 gives the input-output frequency response model of the proposed WEC system
\begin{eqnarray} 
\label{FIO}
X(j\omega)= H(j\omega)F_w(j\omega). 
\end{eqnarray}
{

As seen from \eqref{hydrodynamic_forces},  \eqref{pto force}, \eqref{hydrodynamic_A} and \eqref{FIO}, the proposed WEC is a closed loop system, with a single input $f_w(t)$ ($F_w(j\omega)$ in the frequency domain), 
a single output $x(t)$ ($X(j\omega)$), and the closed loop frequency response function $H(j\omega)$.  As a single-input-single-output system, $f_w(t)$  ($F_w(j\omega)$) is the only source of (mechanical) energy input. The feedback loop is closed by the feedback force $f_{pto}(t)=f_{ei}(t)= f_C(t)+f_R(t)+f_L(t)$, with all three sub-forces 
internally generated by the PMLG and $RLC$ network using only the mechanical input energy. 
Hence, the energy of the output $x(t)$ ($X(j\omega)$) is always less than or equal to the energy of the input $f_w(t)$ ($F_w(j\omega)$). As the input energy from $f_w(t)$ ($F_w(j\omega)$) is always bounded (finite), so is the output energy of $x(t)$ ($X(j\omega)$).  
By signals and systems theory \cite{kwakernaak1991modern}, \eqref{hydrodynamic_A} (\eqref{FIO}) is a passive linear system with BIBO stability.

\textit{\bf Remark 1}: The force $f_{ei}(t)=f_{pto}(t)$ given in \eqref{pto force} is in the same form of the general linear PTO (GLPTO) force,  $f_{pto}(t) = M_{pto} \ddot x(t) + B_{pto} \dot x(t) + K_{pto} x(t)$, implicitly and conceptually proposed in 
\cite{genest2014effect,xu2021maximum}. However, the GLPTO has been utilized mainly in models where the intent is 
maximizing the mechanical power absorption of WECs from ocean waves, but not for maximal electrical power generation. 
As a result, the conceptual GLPTO generally involves additional mechanical and electrical devices that are not used for electric power generation 
 and need external power injection (negative power flow) for their operation. Also, the GLPTO is mostly conceptual. To the best of our knowledge, there has been no 
 physical realization of GLPTO with all the three sub-forces so far; see the references cited in Section I. 
In contrast, the force $f_{ei}(t)=f_{pto}(t)$ in \eqref{pto force} is generated completely  during energy conversion by the energy conversion device comprising the PMLG and $RLC$ network. It is entirely passive without external power injection (without negative power flow) and any additional mechanical devices. Hence, it can be physically and easily implemented. As far as we are aware of, \eqref{hydrodynamic_A} and \eqref{FTF}-\eqref{FIO} are the first of their kind that analytically describes the closed loop hydrodynamics of a whole WEC system, with the PTO feedback gains, $M_{ei}$, $B_{ei}$ and $K_{ei}$, completely determined by the physical parameters of the PMLG and $RLC$, and with the force $f_{pto}(t)=f_C(t)+f_R(t)+f_L(t)$ generated fully internally within the WEC.
\vspace{-0.5em}
\section{Electric output power maximization}
\label{Sect 3}

Applying the inverse Fourier transform to \eqref{FTF}-\eqref{FIO}, it can be readily shown that for 
$f_w(t) = A_w\: \cos(\omega t)$ as given in \eqref{Fe}, the (steady state) output of system \eqref{hydrodynamic_A} is given by 
\begin{equation}
\label{xcos}
    x(t) = |H(j\omega)|A_w \cos(\omega t + \varphi),
\end{equation}
\begin{equation} 
\label{|H|}
|H(j\omega)|=\frac{1}{|(K_m+ K_{ei})-\omega^2(M_m+M_{ei}) + j\omega (B_m+B_{ei})|},
\end{equation} 
\begin{equation}
\varphi = \tan^{-1} \left(\frac{-\omega (B_m+B_{ei})}{(K_m+ K_{ei})-\omega^2(M_m+M_{ei})} \right), 
\end{equation}
where $|H(j\omega)|$ and $\varphi$ are respectively the magnitude and phase of $H(j\omega)$. It then follows from \eqref{xcos} and $-\sin(\omega t +\varphi) = \cos(\omega t +\pi/2+\varphi)$ that
\begin{equation}
\label{x'cos}
    \dot x(t) 
    = \omega|H(j\omega)|A_w \cos(\omega t + \pi/2+\varphi),
\end{equation}
\begin{equation}
\label{accx}
    \ddot x(t) = -\omega^2|H(j\omega)|A_w \cos(\omega t + \varphi).
\end{equation}


As seen from \eqref{e}-\eqref{RLC} and \eqref{xcos}-\eqref{accx}, the amplitudes of internal voltage  $v(t)$, current $i(t)$ and the instantaneous electric power 
$p(t)=v(t)i(t)$
 are all maximized when the amplitude of $x(t)$, $|H(j\omega)|A_w$,
 is maximized. Therefore, maximizing $|H(j\omega)|$ maximizes the amplitude of $x(t)$, which in turn maximizes the amplitude of electric power $p(t)$. 
\vspace{-0.2cm}
\subsection{Maximizing $|H(j\omega)|$ by LC tuning} 
\label{Hmax}
As seen from \eqref{|H|}, $|H(j\omega)|$ is maximized when its denominator 
$|(K_m+ K_{ei})-\omega^2(M_m+M_{ei}) + j\omega (B_m+B_{ei})| =
\{[(K_m+ K_{ei})-\omega^2(M_m+M_{ei})]^2 + [\omega (B_m+B_{ei})]^2\}^{1/2}$
is minimized. From \eqref{Frad} and \eqref{MBK}, $B_m>0$ and $B_{ei}> 0$ for all $0< R < \infty$. Hence, $[\omega (B_m+B_{ei})]^2 > [\omega B_m]^2 > 0$ always holds as long as there is an electric load $0< R < \infty$. 
The only means to minimize $|H(j\omega)|$ under a given $0<R<\infty$ is to tune $K_{ei}$ and $M_{ei}$ through $L$ and $C$ such that, ideally,   
$(K_m+ K_{ei})-\omega^2(M_m+M_{ei})=0,$     
or equivalently
\begin{equation}
\label{0IM}
    \sqrt{(K_m+K_{ei})/(M_m  + M_{ei})}=\omega.
\end{equation}
When \eqref{0IM} is satisfied, 
$H(j\omega) = \frac{1}{j\omega(B_m+B_{ei})}$,
which is purely imaginary with phase $\varphi=-\pi/2$ and maximum magnitude
\begin{equation}
\label{min|H|}
|H(j\omega)|= |H(j\omega)|_{max} := \frac{1}{\omega(B_m+B_{ei})}, 
\end{equation} 
and the system \eqref{hydrodynamic_A} resonates with its input force $f_w(t)=A_w \cos(\omega t)$; 
the amplitudes of $x(t)$, $\dot x(t)$, $\ddot x(t)$, $v(t)$,  $i(t)$ and $p(t)$ are all maximized for the given electric load $R$. 

When $C$ and $L$ are disconnected, $C=0$, $L=\infty$, $M_{ei} = 0$, $K_{ei} = 0$, and  \eqref{0IM} reduces to
\begin{equation}
    \sqrt{K_m/M_m} = \omega_0.
    \label{NRes}
\end{equation}
This $\omega_0$ is the natural frequency (or undamped resonant frequency) of conventional  WECs studied in most of the previous, mechanically-focused, literature 
\cite{manasseh2021fluid}. It is determined by the WECs' inherent mass $M_m$ and stiffness $K_m$. Although, as noted in Section I, many ingenious mechanical techniques have been proposed to actively tune $K_m$ or $M_m$\cite{wang2018review}, the vulnerability of complex mechanical linkages, valves or other devices needed for mechanical tuning has meant that in practice, to the best of our knowledge, $K_m$ and $M_m$ have effectively been fixed for a conventional WEC. When  $\omega \neq \omega_0$ due to changes in  wave frequency, 
the conventional WEC will lose resonance, resulting in lower amplitudes of $x(t)$,  $\dot x(t)$, $\ddot x(t)$, $e(t)$, $i(t)$ and $p(t)$.

Since $M_{ei}$ and $K_{ei}$ (defined in \eqref{MBK}) can be varied by varying $C$ and $L$, the proposed WEC system can be tuned using $C$ and $L$ by the following three Rules derived from  \eqref{0IM} and \eqref{NRes}. 

\begin{enumerate}
\item When $\omega =\omega_0$, 
disconnect $C$ and $L$ to make $C = 0$ and $L = \infty$ in \eqref{MBK} and hence $M_{ei} = 0$ and $K_{ei} = 0$ in \eqref{0IM}. 
\item When $\omega < \omega_0$, disconnect $L$ to make $K_{ei} = 0$ and tune $C$ to make $M_{ei} = K_m/\omega^2 -M_m$
in \eqref{0IM}. From \eqref{MBK}, the $C$ (in Farad, F) thus tuned  is given by 
\begin{equation} 
\label{Cvalu}
C= \frac{M_{ei}}{K_tK_e} = \frac{K_m/\omega^2-M_m}{K_tK_e} > 0.
\end{equation}
\item When  $\omega > \omega_0$, 
disconnect $C$ to make  $M_{ei} = 0$ and tune $L$ to make $K_{ei} =\omega^2M_m-K_m$
in \eqref{0IM}. From \eqref{MBK}, the $L$ (in Henry, H) thus tuned is given by
\begin{equation}
\label{Lvalu}
    L= \frac{K_tK_e}{K_{ei}} = \frac{K_tK_e}{\omega^2M_m-K_m} >0.    
\end{equation}
\end{enumerate}

These three Rules yield the same $H(j\omega)$ with $|H(j\omega)|=|H(j\omega)|_{max} = \frac{1}{\omega(B_m+B_{ei})}$
and  phase 
$\varphi=-\pi/2$ for all $\omega >0$. 
It then follows from \eqref{x'cos}
that under Rules 1)$-$3) 
\begin{equation}
\label{x'r}
\dot x(t) 
\equiv \frac{A_w}{B_m+B_{ei}}\cos(\omega t), \forall \omega >0,
\end{equation}
which is in phase with $f_w(t)=A_w\cos(\omega t), \forall \omega >0$ and gives the instantaneous mechanical absorption of WEC \begin{equation}p_{a}(t) := f_w(t)\dot x(t) = \frac{A_w^2}{B_m+B_{ei}}\cos^2(\omega t)\geq 0, \forall t \label{pa}.
\end{equation}
Therefore, the system \eqref{hydrodynamic_A} always resonates with the input force $f_w(t)$ and the amplitudes of $x(t)$, $\dot x(t)$, $\ddot x(t)$, $v(t)$,  $i(t)$ and $p(t)$ are always maximized for the given electric load $R$, irrespective of the wave frequency $\omega$.

It can be seen from \eqref{fc}-\eqref{fl}, \eqref{xcos}, \eqref{x'cos}, \eqref{accx} and \eqref{x'r} that with Rules 1)$-$3), $f_R(t)=B_{ei}\dot x(t)$ is in phase with $f_w(t)$; $f_C(t) = M_{ei} \ddot x (t)$ leads 
and $f_L(t) = K_{ei} x(t)$ 
lags $f_w(t)$ by $\pi/2$ rad. As shown in \ref{MP}, only $f_R(t)$ absorbs average mechanical power. Hence, they are respectively the active, reactive leading and reactive lagging PTO sub-forces. 
They together produce the active and reactive PTO forces for  different wave frequencies under Rules 1)$-$3), as shown in \ref{EP}.  
Moreover, from \eqref{Cvalu} and \eqref{Lvalu}, the lower the $\omega < \omega_0$, the larger the $C$ is required; and the higher the $\omega > \omega_0$, the smaller the $L$ is required.
\vspace{-1em}
\subsection{Maximizing mechanical power absorption by optimal $R$}
\label{MP}
From \eqref{hydrodynamic_A},
the average mechanical power  absorption $P_{a}$ of the resonating WEC, over a full period $T$ of $f_w(t)$, 
can be written as 
\begin{eqnarray}
P_{a} &=& \frac{1}{T}\int_T p_{a}(t) dt = \frac{1}{T}\int_Tf_w(t)\dot x(t) dt \nonumber\\ 
&= &\frac{1}{T}\int_T\left[M_m\ddot x(t) + B_m \dot x(t) + K_m x(t)\right]\dot x(t) dt +  \label{P3} \nonumber\\
& &\frac{1}{T}\int_T \left[ M_{ei}\ddot x(t) + B_{ei} \dot x(t) + K_{ei} x(t)\right]\dot x(t) dt\\ 
&= &\frac{B_m}{T}\int_T \dot x^2(t) dt + \frac{B_{ei}}{T}\int_T\dot x^2(t) dt \label{PB} \\
&=& \frac{A_w^2B_m}{2(B_m+B_{ei})^2} + \frac{A_w^2B_{ei}}{2(B_m+B_{ei})^2}\label{PBM} \\ 
&=:& P_m + P_{ei}. \label{PBM2} 
\end{eqnarray}
In the above, \eqref{PB} follows from \eqref{xcos}, \eqref{x'cos} and \eqref{accx} which show that $\ddot x(t)$ and $x(t)$  are orthogonal to $\dot x(t)$, when the WEC resonates at the frequency $\omega$  with $\varphi = -\pi/2$,
and hence the first and third terms in the two integrals of \eqref{P3} integrate to zero; 
\eqref{PBM} follows from substituting \eqref{x'r} into \eqref{PB} and using $\cos^2(\omega t) = 1/2 + \cos(2\omega t)/2$. 
As seen  from 
\eqref{P3}-\eqref{PBM2}, $P_{ei}$ and $P_m$ are the average mechanical powers absorbed by WEC through the 
PMLG induced damping force $B_{ei}\dot x(t)$  and the wave induced damping (radiation) force $B_m \dot x(t)$. 
Here $B_m$ is the unchangeable inherent mechanical damping of WEC, while $B_{ei}=K_tK_e/R$ can be tuned using $R$ to maximize $P_{m}$, $P_{ei}$ and hence $P_{a}$.

Taking $dP_{ei}/dB_{ei}  = A_w^2 \frac{(B_m+B_{ei}) - 2B_{ei}}{2(B_r+B_{ei})^3} = 0$ and solving for $B_{ei}$ gives the optimal $B_{ei}$ and the corresponding optimal electric load $R$, denoted as $B_{ei}^*$ and $R^*$ (in Ohm),
\begin{equation}
\label{optB}
B_{ei}^*  = B_m = K_tK_e/R^*, \quad R^* = K_tK_e/B_m,
\end{equation}
and the resulting maximal average mechanical absorption $P_{ei}$, $P_m$ and $P_a$, denoted as $P_{ei}^*$, $P_m^*$ and $P_a^*$ (in Watt), respectively.
\begin{equation}
\label{optP}
P_{ei}^* = P^*_m = A_w^2/8B_m,  \quad 
P^*_a = 2P_{ei}^* = 2P^*_m= A_w^2/4B_m. 
\end{equation}
Note that \eqref{optB} and \eqref{optP} hold for all $\omega >0$ due to Rules 1)$-$3).

\textit{\bf Remark 2}: It is clear from \eqref{P3} and \eqref{PB} that the average mechanical power absorbed by the reactive sub-forces $(M_m+M_{ei})\ddot x(t)$ and  
$(K_m+K_{ei})x(t)$ is zero. This and the maximal average power absorption \eqref{optP} 
are consistent with previous results, eg \cite{ringwood2014energy,manasseh2021fluid,park2016active}, but they now hold for all $\omega >0$, instead of $\omega = \omega_0$ only,
due to Rules 1)$-$3). The optimal electric load $R^*$ 
and its maximal average power absorption $P_{ei}^*$ are the first of their kind to the best of our knowledge.

\vspace{-0.2cm}
\subsection{Electric active power generated and reactive power needed}
\label{EP}
From \eqref{e} and \eqref{x'r}, the output voltage $v(t)$ under Rules 1)$-$3) has the same amplitude as given below for all $\omega>0$.
\begin{equation}
\label{ecos}
    v(t)=K_e\dot x(t)=\frac{K_eA_w}{B_m+B_{ei}}\cos(\omega t)=\sqrt{2}V \cos(\omega t),    
\end{equation}
where
$V =\sqrt{\frac{1}{T} \int_T v^2(t) dt}= K_eA_w/\sqrt{2}(B_m+B_{ei})$, in Volt, 
is the root mean square (RMS) of $v(t)$.
It then follows from basic AC circuit analysis \cite{nilsson2008electric} that 
the current $i(t)$ is given by
\begin{equation}
\label{it}
    i(t) = \sqrt{2}I \cos(\omega t + \psi),
\end{equation}
where 
$I = V \sqrt{(1/R)^2+(\omega C-1/\omega L)^2}$, in Ampere, 
is the RMS of $i(t)$ and $\psi$ is the phase of $i(t)$ vs $v(t)$ given by 
\begin{eqnarray}
\label{psi}
\psi &=&\tan ^{-1}[(\omega C-1/\omega L)/(1/R)] \nonumber\\ 
&= &\cos^{-1}\left((1/R) / \sqrt{(1/R)^2+(\omega C-1/\omega L)^2} \right).
\end{eqnarray}
The instantaneous electric power can therefore be written as
\begin{eqnarray}
p(t) &=&  v(t)i(t) = 2VI \cos(\omega t)\cos(\omega t + \psi) \nonumber\\ 
&=& VI\cos(\psi) + VI\cos(2\omega t+ \psi), \label{p}
\end{eqnarray}
and the average electric output power can be written as
\begin{eqnarray}
P &=& \frac{1}{T}\int_T p(t) dt 
= \frac{1}{T}\int_T VI[\cos(\psi) + \cos(2\omega t+ \psi)] dt \nonumber\\ 
&=& VI \cos(\psi) = S\cos(\psi), \label{P}
\end{eqnarray}
where $P$ (in Watt) and $S := VI$ (in VA) are respectively the active and apparent electric power outputs of the PMLG, and 
\begin{eqnarray}
\label{PF}
   \cos(\psi) = (1/R)/\sqrt{(1/R)^2 + (\omega C - 1/\omega L)^2} := PF,
\end{eqnarray}
derived from \eqref{psi}, is the power factor ($PF$) of the PMLG. 

When $\omega = \omega_0$ and Rule 1) is used, $\psi = 0$, $i(t) = i_R(t) = v(t)/R$ is in phase with  $v(t)$, and
$PF = 1$. When $\omega < \omega_0$ and Rule 2) is used, $\psi > 0$, $i(t)= i_R(t)+i_C(t)$ leads $v(t)$, and $ PF <1$. When $\omega > \omega_0$ and Rule 3) is used, $\psi < 0$, $i(t)= i_R(t)+i_L(t)$ lags $v(t)$, and $PF <1$. Recall that $v(t)=K_e\dot x(t)$ and $\dot x(t)$ is in phase with $f_w(t)$ as shown in \eqref{x'r}.  Hence, under Rule 1), $f_{pto}(t) = K_ti_R(t) = f_R(t)$ yields the active PTO force in phase with $f_w(t)$; under Rule 2),  $f_{pto}(t) = K_ti_R(t) + K_t i_C(t) = f_R(t) + f_C(t)$ yields the reactive PTO force leading $f_w(t)$;
under Rule 3), $f_{pto}(t) = K_ti_R(t) + K_t i_L(t) = f_R(t) + f_L(t)$ yields the reactive PTO force lagging $f_w(t)$.



With $V$, $I$ and $\cos(\psi)$ defined respectively in \eqref{ecos}, \eqref{it} and \eqref{PF}, the active electric power $P$ can be rewritten as
\begin{eqnarray}
    P 
    &=& VI \cos(\psi) = V^2\cos(\psi)\sqrt{(1/R)^2+(\omega C-1/\omega L)^2} \label{VV} \nonumber\\
    &=& V^2/R 
    = K_e^2A_w^2/2R(B_m+B_{ei})^2, \label{PKe}
\end{eqnarray}
which is dependent on $R$ but independent of $PF$. This is because $R$, $L$ and $C$ are in parallel connection with the same voltage source $v(t)$ and only $R$ absorbs the active power $P$. 

From \eqref{e}, \eqref{f}, 
\eqref{p}, \eqref{P} and $K_e = K_t$, it follows that for any 
$0<R<\infty$,
$p(t)=v(t)i(t)=(K_e/K_t)f_{ei} (t)\dot x(t)= f_{ei} (t)\dot x(t)= p_{ei}(t)$
and
$P = \frac{1}{T}\int_T p(t) dt = \frac{1}{T}\int_T p_{ei} (t) dt 
= P_{ei}$. 
By \eqref{optB}, \eqref{optP} and \eqref{PKe},  when $R=R^*=K_tK_e/B_m$, 
$P_{ei} = P^*_{ei}=A_w^2/8B_m$ for all $\omega>0$, which is the maximal average mechanical power the PMLG can absorb.  
Hence, when the electric load $R=R^*$, the maximal active electric power output $P^*=P^*_{ei} = A_w^2/8B_m$  W is attained for all $\omega >0$. 

Although $P$ 
is independent of $PF$, the apparent power $S= VI = P/PF$ does increase as $PF$ decreases. 
Physically, $PF$ decreases when $\omega C$ or $1/\omega L$ increases, which increases $i_C(t)$ or $i_L(t)$ and hence $i(t) =i_R(t)+i_L(t)+i_C(t)$ and its RMS (aka effective) value $I = V \sqrt{(1/R)^2+(\omega C-1/\omega L)^2}$.
The increased $i(t)$ in turn increases the instantaneous PTO force $f_{pto}(t) 
=  K_t i(t)$ and the effective PTO force $F_{pto} 
= K_t I$ to keep resonance at the frequency $\omega \neq \omega_0$. 
Consequently, the reactive power output, given by $Q=S\sin(\psi)$ (in VAR), increases. When $PF$ is low, the  efficiency of the PMLG in active electric power generation becomes low. 
For the same level of $P$ output, the PMLG's capacity rating $S = VI$ must be higher at lower $PF$ since $I$ flowing through the PMLG is higher.
Therefore, to attain resonance and hence maximal active electric  power output over a wider range of wave frequencies, we must use a PMLG with higher capacity (VA) rating to withstand higher $I$, and capacitor $C$ with larger capacitance to provide higher $I$. This is the price we have to pay. 
\vspace{-1em}

\section{Simulation Studies} 
}
%

Simulation studies are conducted to verify the proposed WEC 
and Tuning Rules 1)$-$3)  and compare with the WEC without $LC$ tuning. 
Simulations are performed using continuous-time dynamic system modules and the \verb"ode15s" solver of MATLAB/Simulink with a time step of 0.01 seconds. The simulated WEC, having a buoy of radius $b_r$ = 1.5 m and height $b_h$ = 2.75 m, is adopted from \cite{sousounis2017direct} with $M_m$ = 10,000 kg, $B_m$ = 4,000 Ns/m, $K_m$ = 31,580 N/m, $\omega_0$ = $\sqrt{K_m/B_m} = 1.77$ rad/s, $K_e = 842$ Vs/m, $K_t = 842$ N/A, and poles $N_p = 2$. The simulated wave input force $f_w(t)= A_w \cos(\omega t)$ has $A_w = 10$ kN and its frequency $\omega$ varied in the studies presented below.


From (\ref{optB})-(\ref{optP}) and the analysis of Section III, the optimal electric resistor load  for the simulated WEC is $R^* = K_e K_t/B_m = 177$ $\Omega$, which renders the maximum average mechanical power absorption $P^*_a $ $= A_w^2/4B_m = 6.25$ kW 
and induces the maximum 
active electric output power 
$P^* = P^*_a/2 = 3.125$ kW. Also, as given in (\ref{x'r}), with $R = R^* = 177.5$ $\Omega$ and Rules 1)$-$3), the velocity of the WEC,  $\dot x(t)$,
is in phase with the input force $f_w(t)$ 
for all $\omega >0$.  Hence, we set $R = 177$ $\Omega$, and use these $P^*_a$ and $P^*$ values
and $\dot x(t)$ in phase with $f_w(t)$ as the theoretical values and condition to check the simulation results in all the studies. 

To facilitate comparison, we use blue curves to show the waveformes and superscript $^*$ to represent the variables of tuned WEC, 
and use black curves for the waveformes of untuned WEC and red curves for the input force. The curves in Cases 1-3 are plotted from static start to steady state operation.
\\

\vspace{-0.2cm}
\subsection{Case 1 ($\omega = \omega_0 = 1.77$ rad/s)}
By Rule 1), the proposed WEC in this ideal case does not need tuning and the $LC$ network is disconnected.
Figs.~\ref{resonance_fvd}-\ref{resonance_power} show the waveforms of the relevant variables of the WEC,   
which approach respective steady state amplitudes in about 7 s.
As analyzed in Section \ref{Sect 3}, the WEC resonates at the frequency $\omega = \omega_0 = 1.77$ rad/s, with its $\dot x(t)$ ($v(t) = K_e \dot x(t)$) and $f_{pto}(t)$ ($i(t) = f_{pto}(t)/K_t$) all in phase with $f_w(t)$; the amplitude of $f_{pto}(t)$ is well below that of $f_w(t)$; 
it absorbs the maximal mechanical power $P_a=P_a^* = 6.25$ kW, and
the PMLG outputs the maximal active electric power $P=P^* = 3.125$ kW $= P^*_{pto}$, the maximal mechanical power it can absorb, with $PF=1$. The PMLG capacity rating in this case must be $\geq$ $S^* = P^* =3.1255$ KVA. 
\begin{figure}[t!]
  \centering  
  \includegraphics[width= 8.5cm, height = 4.5cm ]{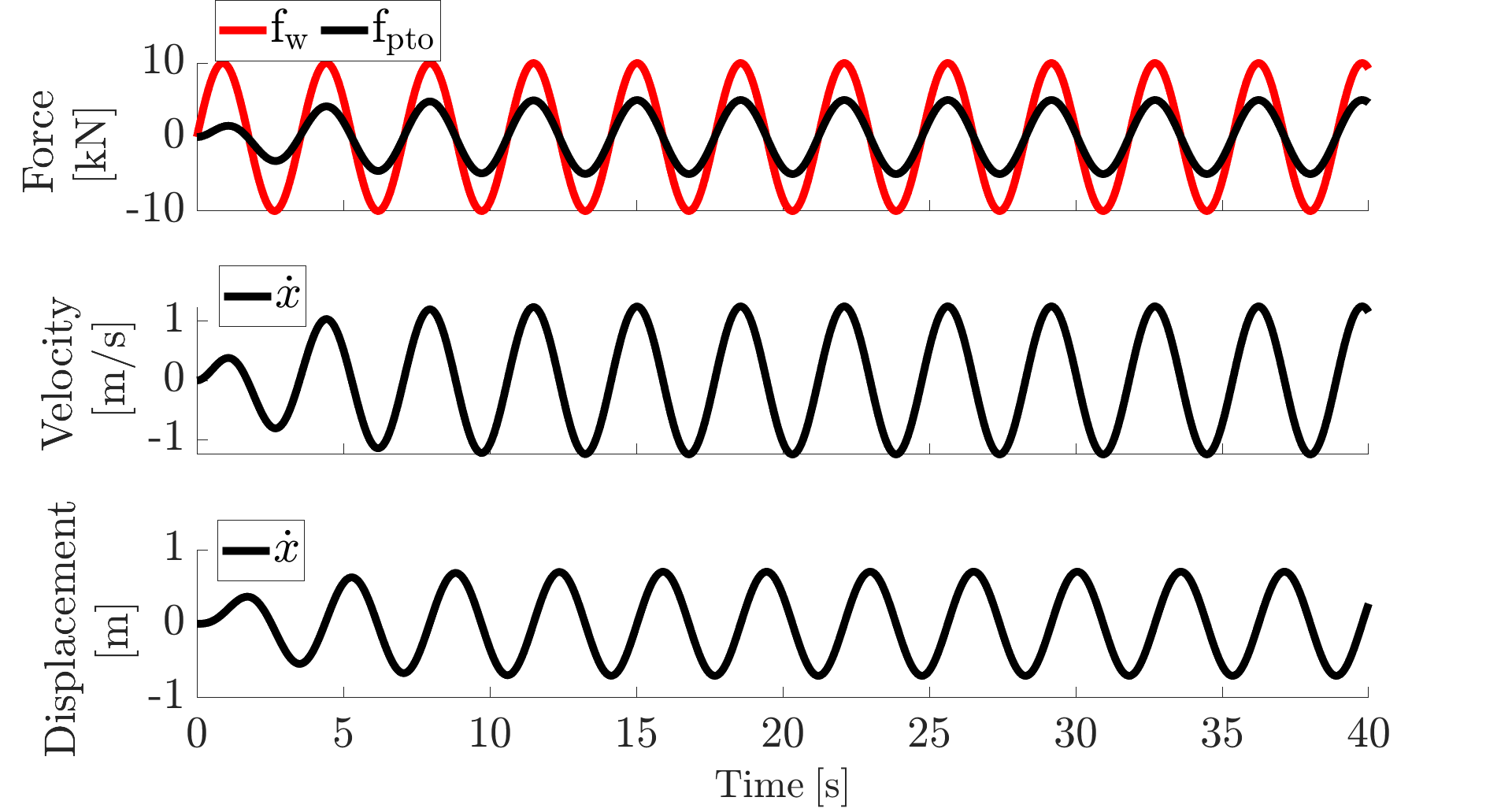}
  \caption{WEC response at natural resonance frequency $\omega = \omega_0$ 
  }
  \label{resonance_fvd}
\end{figure}
\vspace{-0.4cm}
\begin{figure}[]
  \centering
\includegraphics[width= 8.5cm, height = 4.5cm ]{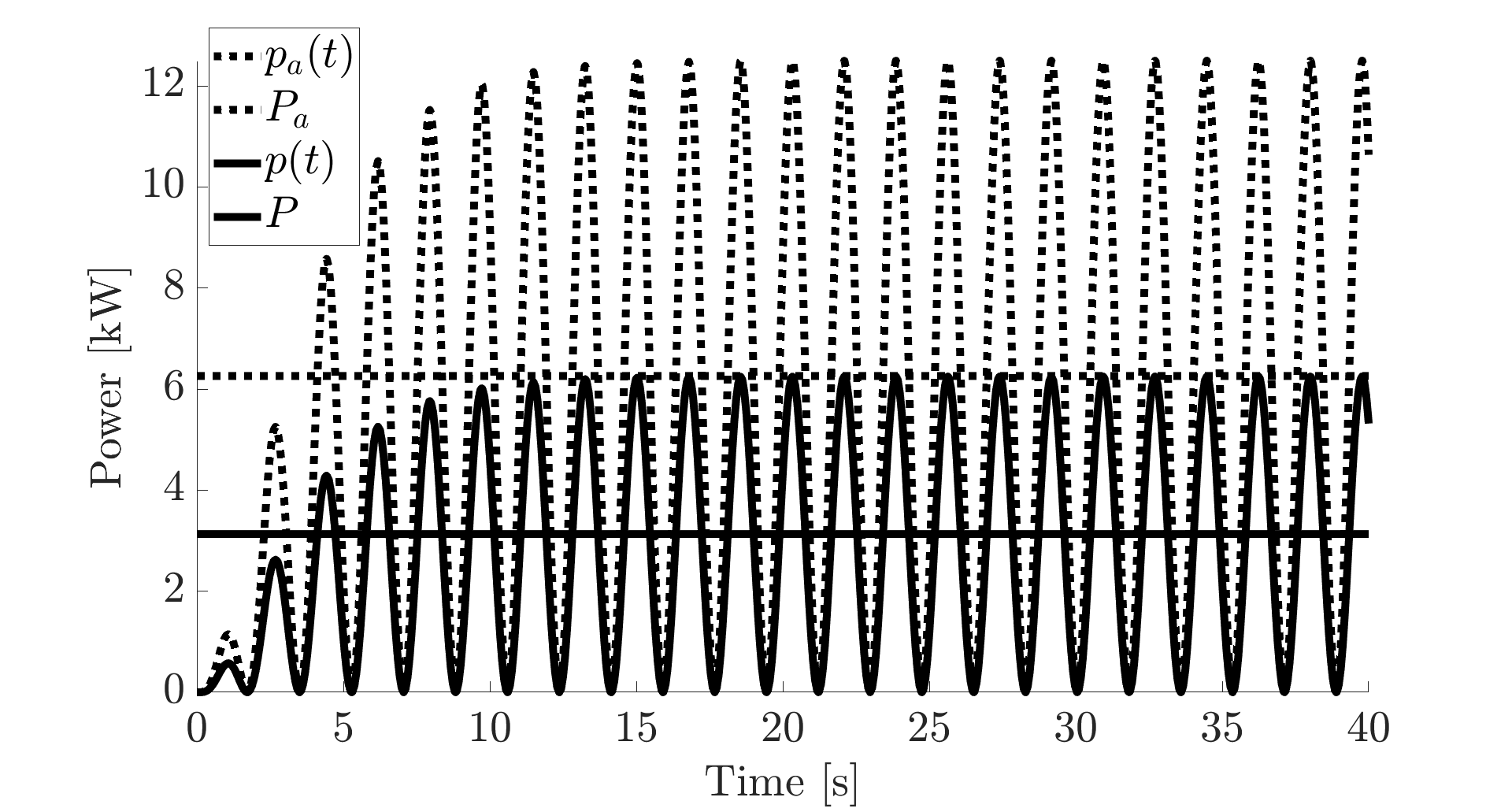}
\caption{Mechanical and active electrical powers of WEC at $\omega = \omega_0$ }
  \label{resonance_power}
\end{figure}

\subsection{Case 2 ($\omega = 1$ rad/s $< \omega_0 = 1.77$ rad/s)}
By Rule 2), the proposed WEC in this case is tuned with $C = 0.0304$ $F$ and $L$ is disconnected.
Figs.~\ref{1fvd}-\ref{1c_power} show the waveforms of the relevant variables of tuned and untuned WECs.
The tuned and untuned WECs both oscillate at the  frequency of $f_w(t)$, $\omega = 1$ rad/s, and the tuned waveforms 
approach respective steady state amplitudes in about 30 s. Compared with those of Case 1, the steady state amplitudes of tuned waveforms are the same or higher, whereas those of untuned waveforms are much lower. This is because the untuned WEC is not resonating with $f_w(t)$. Hence, its average mechanical power absorption $P_a = 1$ kW and active electric power output $P = 0.5$ kW, as opposed to $P_a^* = 6.25$ kW and $P^* = 3.125$ kW of the tuned WEC.

For the untuned WEC, the phases of its $f_{pto}(t)$, $\dot x(t)$ and $x(t)$ lead their tuned counterparts.
For the tuned WEC, the phases of its  $\dot x^*(t)$ and $x^*(t)$ remain the same as those in Case 1, with $\dot x^*(t)$ still in phase with $f_w(t)$, but its $f_{pto}^*(t)$ 
is no longer in phase with $f_w(t)$ as was in Case 1, instead, it now leads $f_w(t)$. 
The tuning capacitor $C$ 
introduces $i_C(t)$ into $i^*(t) = i_R(t) + i_C(t)$ to significantly boost the amplitudes of $i^*(t)$ and $f_{pto}^*(t) = K_t i^*(t)$ and makes $i^*(t)$ leading $v^*(t) = K_e \dot x^*(t)$ and $f_{pto}^*(t) = K_t i^*(t)$ leading $f_w(t)$ by $\psi = 1.3886$ rad. 
The amplitude of this reactive leading PTO force $f^*_{pto}(t)$ is about 2.2 times higher than $f_w(t)$ and  4 times higher than in Case 1, which
forces the tuned WEC to resonate at $\omega = 1$ rad/s instead of $ \omega_0 = 1.77$ rad/s  and output the maximal active electric power $P^* = 3.125$ kW. 
This is at the cost of low $ PF = \cos(\psi) = 0.18$ and high $S^* ~= P^*/PF = 17.25$ KVA of PMLG  as analyzed in \ref{EP}. So the PMLG capacity rating must be $\geq$ 17.25 KVA in order to maintain the 3.125 kW maximal active electric power output, when the wave frequency $\omega = 1$ rad/s is about $44\%$ lower than WEC's natural frequency $\omega_0 =1.77$ rad/s. 



\begin{figure}[t]
  \centering
  \includegraphics[width= 8.5cm, height = 4.5cm ]{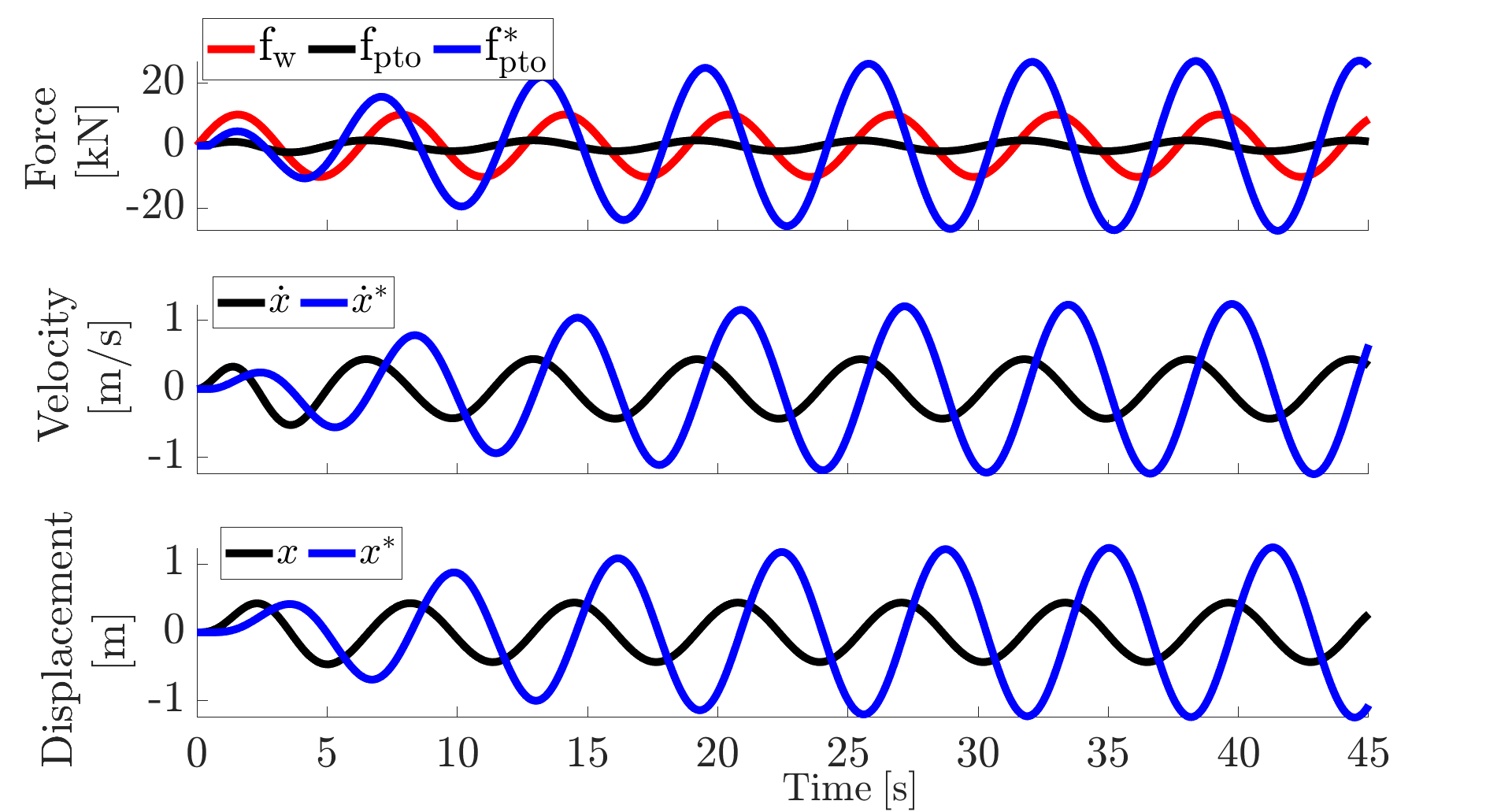}
  \caption{ Responses of tuned and untuned WECs at $\omega < \omega_0$, 
  }
  \label{1fvd}
\end{figure}
\begin{figure}[!]
  \centering
  \includegraphics[width= 8.5cm, height = 4.5cm ]{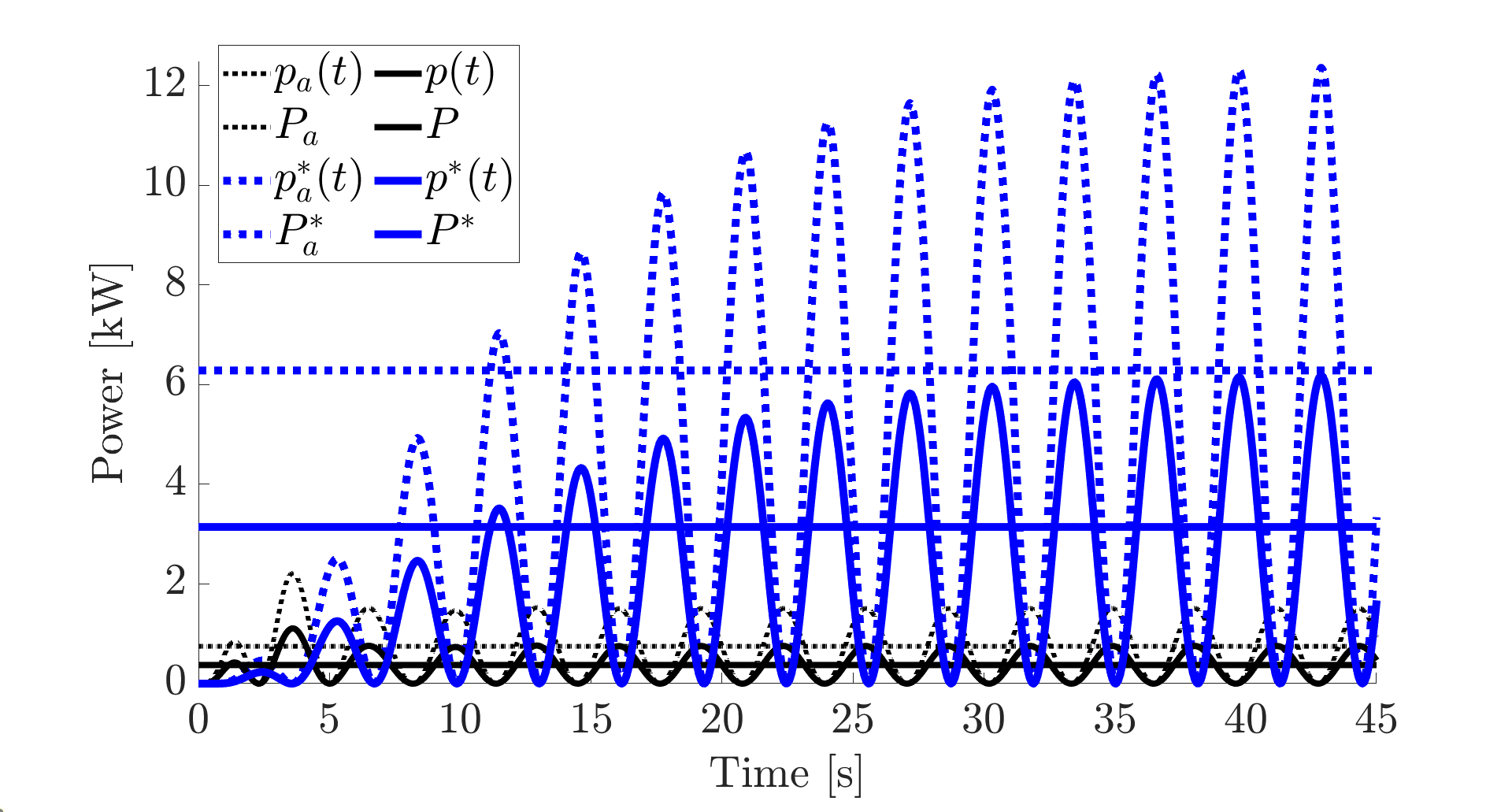}
  \caption{Mechanical and active electrical powers of tuned and untuned WECs at $\omega < \omega_0$ 
  }
  \label{1c_power}
\end{figure}
%
%
%

\subsection{Case 3 ($\omega = 2.3$ rad/s $> \omega_0 = 1.77$ rad/s)}
By Rule 3), the proposed WEC in this case is tuned with C disconnected and $L = 33.3113$ H. 
As seen from  Figs. \ref{2fvd}-\ref{2_power}, similar to Case 2, both WECs oscillate at the frequency of $f_w(t)$, $\omega = 2.3$ rad/s, and the tuned waveforms
approach respective steady state amplitudes in about 12 s. The waveform amplitudes of the untuned WEC are significantly lower because it is not resonating with $f_w(t)$. Its average mechanical power absorption $P_a = 2.6$ kW and its active electric power output $P = 1.3$ kW, as opposed to $P_a^* = 6.25$ kW and $P^* = 3.125$ kW of tuned WEC. 
The phases of $f_{pto}(t)$, $\dot x(t)$ and $x(t)$ of the untuned WEC now lag their tuned counterparts. The phases of $\dot x^*(t)$ and $x^*(t)$ of the tuned WEC remain the same as those in Case 1, 
with $\dot x^*(t)$ still in phase with $f_w(t)$, 
but its $f_{pto}^*(t)$ now lags $f_w(t)$. The tuning inductor $L$ introduces $i_L(t)$ into $i^*(t) = i_R(t) + i_L(t)$ to significantly boost the amplitudes of $i^*(t)$ and $f_{pto}^*(t) = K_t i^*(t)$ and make $i^*(t)$ lagging $v(t) = K_e \dot x^*(t)$ and $f_{pto}^*(t) = K_t i^*(t)$ lagging $f_w(t)$ by $\psi = 1.1674$ rad. 
The amplitude of this lagging reactive $f_{pto}^*(t)$ is about 20\% higher than $f_w(t)$ and 2 times higher than in Case 1, which
forces the tuned WEC to resonate at $\omega = 2.3$ rad/s, instead of $\omega_0 = 1.77$ rad/s, and output $P^* = 3.125$ kW. The cost for this is the low $ PF = \cos(\psi) = 0.3925$ and high $S^* =P^*/PF = 7.92$ KVA of PMLG as analyzed in \ref{EP}. The PMLG capacity rating must be $\geq$ 7.92 KVA in order to maintain the 3.125 kW maximal active electric power output, when the wave frequency $\omega = 2.3$ rad/s is about $30\%$ higher than WEC's natural frequency $\omega_0 =1.77$ rad/s. 

\begin{figure}[t!]
  \centering
 \includegraphics[width= 8.5cm, height = 4.5cm ]{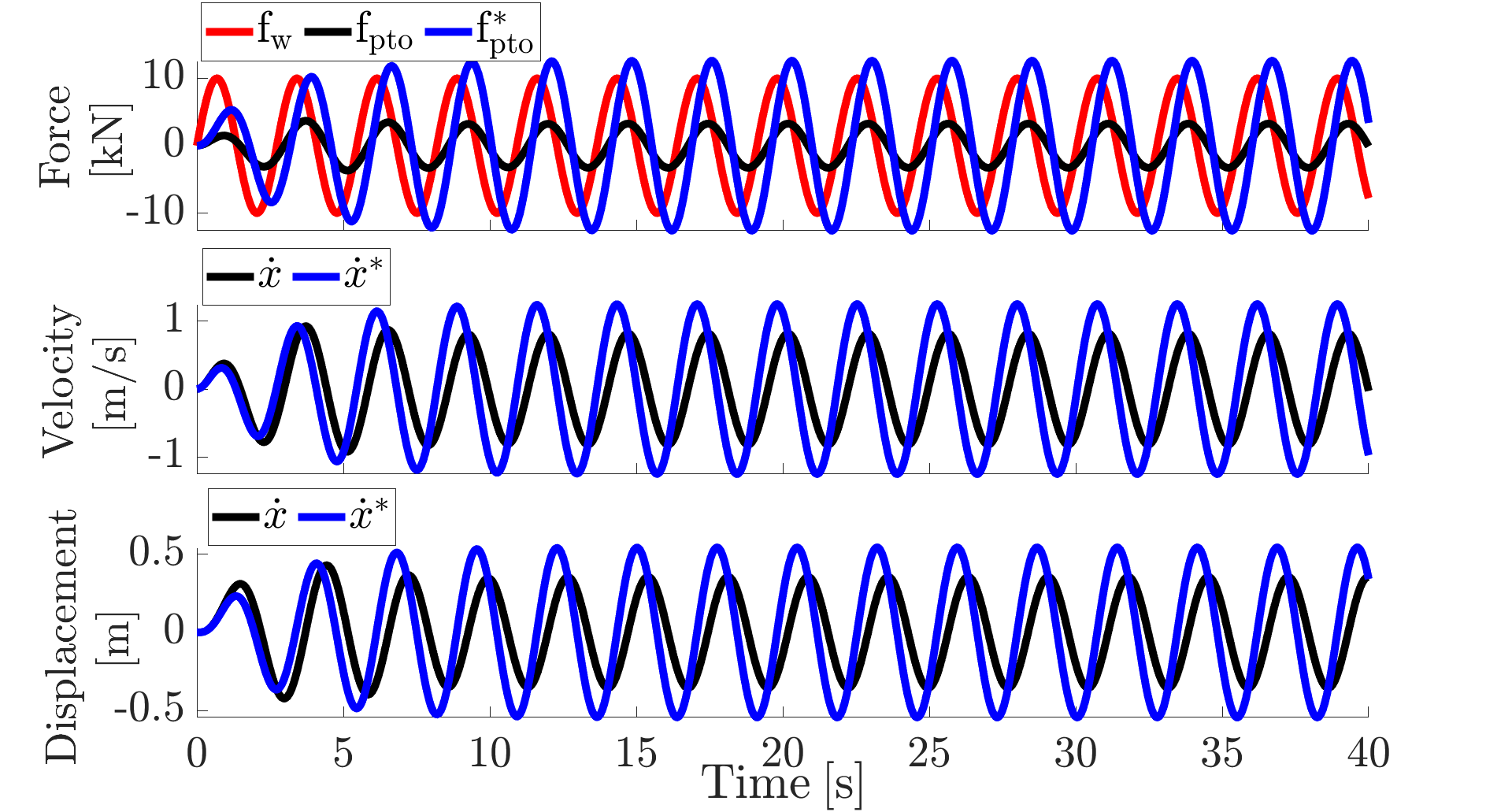}
  \caption{Responses of tuned and untuned WECs at $\omega > \omega_0$. 
  }
  \label{2fvd}
\end{figure}
\begin{figure}[t!]
  \centering
    \includegraphics[width= 8.5cm, height = 4.5cm ]{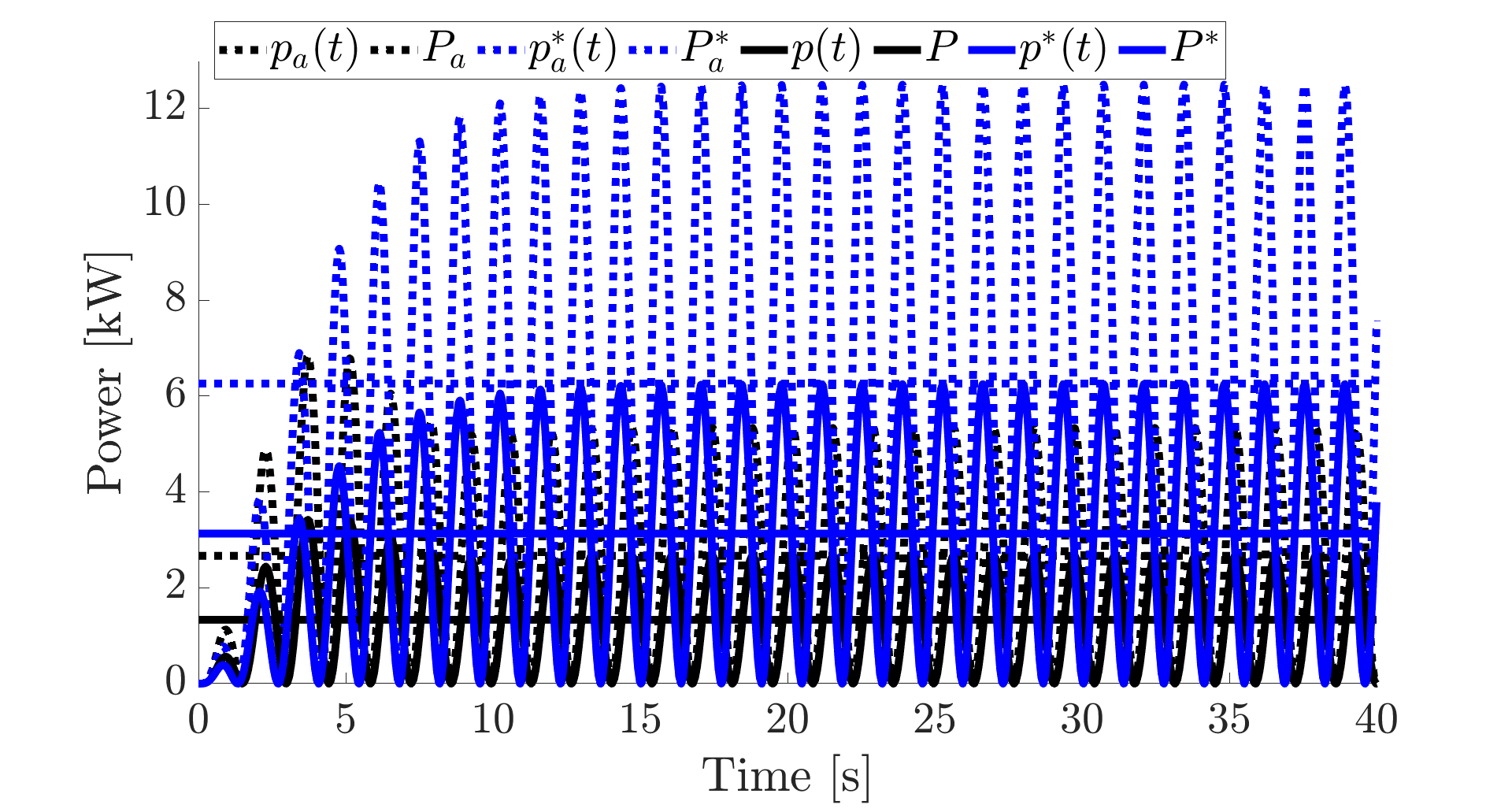}
  \caption{Mechanical and active electrical powers of tuned and untuned WECs at $\omega > \omega_0$ 
  }
  \label{2_power}
\end{figure}
%
\vspace{-0.2cm}
\subsection{Case 4 ($0.5 $ rad/s $ \leq \omega \leq 2.7$ rad/s)}
The proposed WEC in this case is tuned with Rules 1)-3) over $\omega \in [0.5, 2.7]$ rad/s and is compared with the untuned WECs over the same frequencies.  Fig. \ref{apparent} plots  
the effective input force $F_w$ (RMS of $f_w(t)$), the RMS current $I$, effective PTO force $F_{pto} = K_t I$ and apparent power $S = VI$ of the PMLG in the tuned and untuned WECs over $\omega \in [0.5, 2.7]$, where $I$, $F_{pto}$ and $S$ share the same curves but with different scales since $K_t$ and $V$ are constants for all $\omega >0$ as shown earlier. As seen from the plots, for the untuned WEC, its $F_{pto} < F_w, \forall \omega >0$, and its $I$, $F_{pto}$ and $S$ fall off from the peak at $\omega = \omega_0$ when $\omega$ is above or below $\omega_0$ (the black curve of Fig. \ref{apparent}), resulting in the decreased average mechanical power absorption $P_a$ and active electric power output $P$ (the black curves of Fig \ref{mpeak}). In contrast, for the tuned WEC, its $F^*_{pto}$ increases as $\omega$ shifts away from $\omega_0$ and becomes significantly higher than $F_w$ when $\omega$ is significantly lower or higher than $\omega_0$ (the blue curve of Fig \ref{apparent}); this increased $F^*_{pto}$ drives the tuned WEC to resonate with the input force at all $\omega >0$ with significantly large $I^*$ and $S^*$, and keeps its average mechanical power absorption $P_a^*$ and active electric power output $P^*$ constant for all $\omega >0$ (the blue lines of Fig \ref{mpeak}). As shown by the two example blue curves in Fig \ref{mpeak}, Rules 1)-3) effectively shift the power curve as the input force frequency changes and align the power peak with the input force frequency $\omega$. The apparent power $S^*$ shown by the blue curve of Fig. \ref{apparent} gives the lower bound for the PMLG capacity rating required at different wave frequencies.

\begin{figure}[t!]
  \centering
\includegraphics[width= 8.5cm, height = 4.5cm]{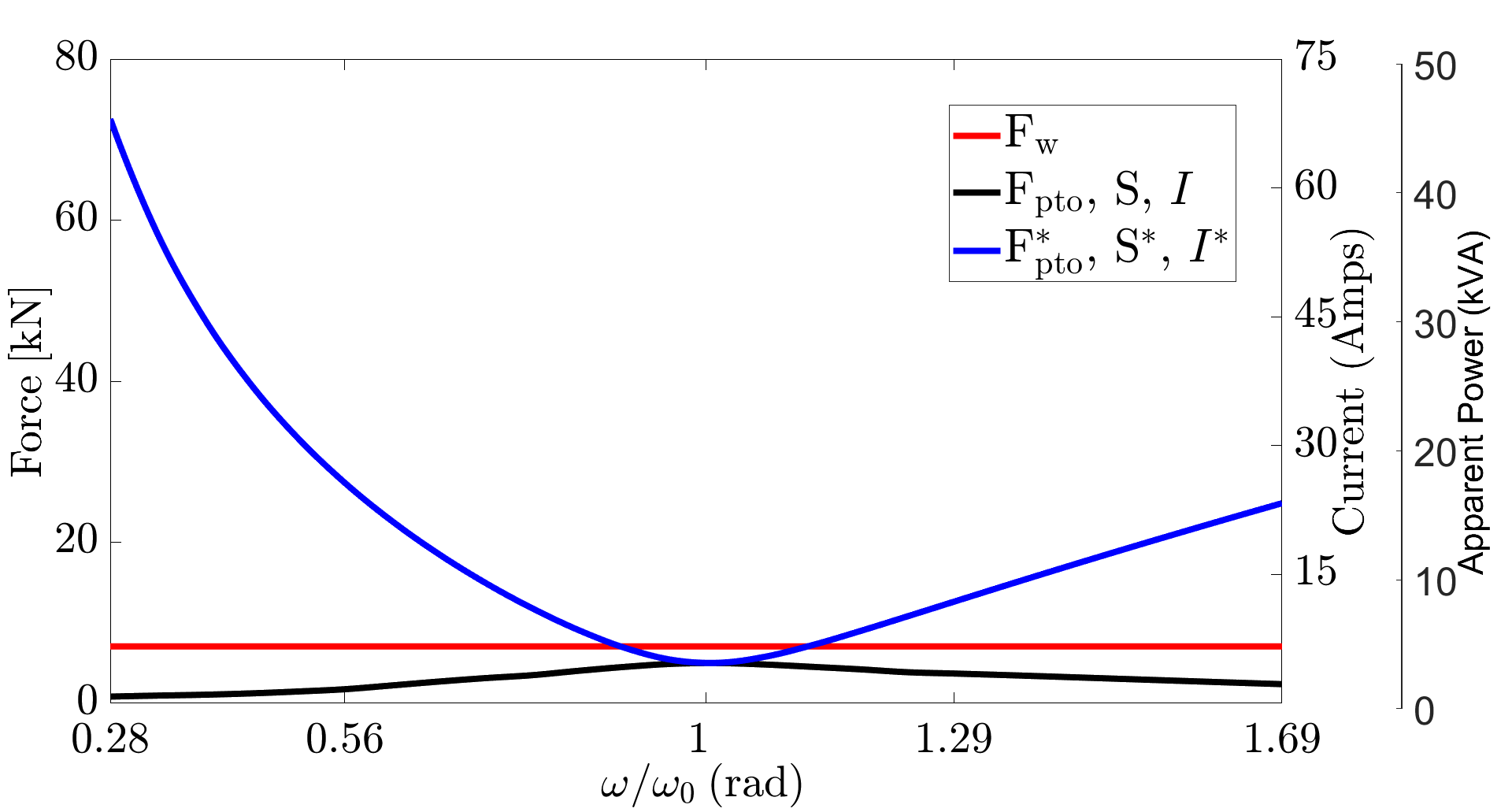}
\caption{RMS Currents $I$, effective PTO forces $F_w$ and apparent powers $S$ of PMLG in tuned and untuned WECs over $\omega \in [0.5, 2.7]$. 
}
\vspace{-1em}
\label{apparent}
\end{figure}
%

\begin{figure}[t!]
  \centering
 \includegraphics[width= 8.5cm, height = 4.5cm] {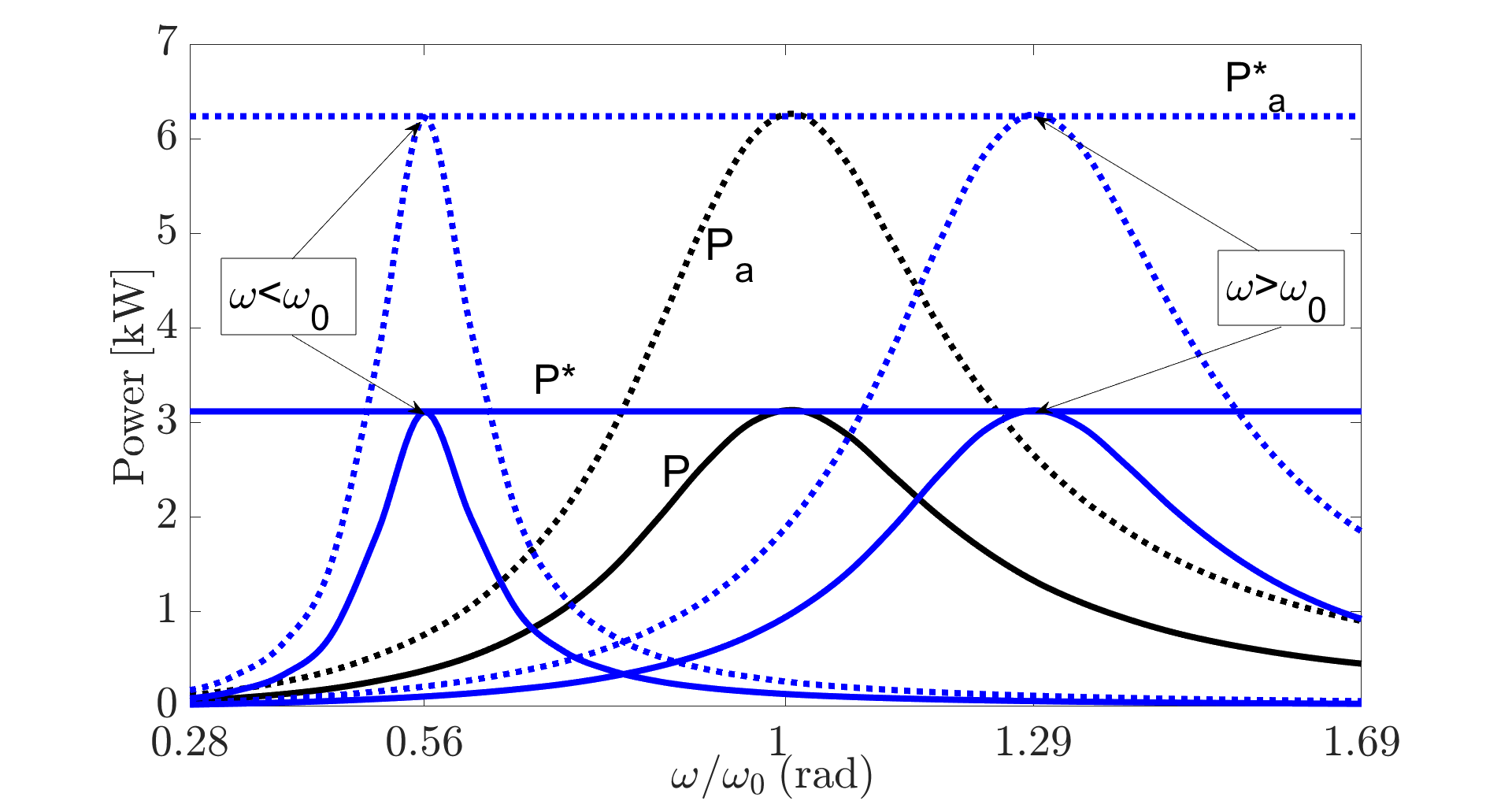}
\caption{Active powers vs wave frequency of tuned (blue) and untuned (black) WECs}
  \vspace{-1em}
  \label{mpeak}
\end{figure}
%
\section{Discussions}
 

Except for complicating the derivation and analysis, the neglected internal resistance $R_{in}$ and inductance $L_{in}$ of the PMLG do not affect the main results given above. 
The $L_{in}$ yields some lagging in $i(t)$ and $f_{pto}(t)=K_ti(t)$, which can be easily compensated for by increasing the value of $C$ if needed; $R_{in}$ consumes some active electric power and reduces the net active electric power output of the PMLG, which is inevitable in all generators. 

As shown analytically and experimentally
in Sections III$-$IV, the capacity of the PMLG must be  $\geq S^* = P^*/PF$ VA, where $P^* = A^*_w/8B_m$W, given in \eqref{optP}, is the maximal active electric power the WEC can produce, and $0< PF < 1$ depends on the frequency range of $f_w(t)$ over which we want WEC to resonate. The leading  (lagging) current $i_C(t)$ ($i_L(t)$) from $C$ ($L$) and the corresponding reactive leading (lagging) $f_{pto}(t)$ are the key to the resonance at $\omega \neq \omega_0$. Limiting $i_C(t)$ ($i_L(t)$) will limit the frequency range of resonance or the oscillation amplitude of the WEC at $\omega \neq \omega_0$, leading to reduced electric power generation. Although these results are obtained for steady state operation of WEC without dynamic PTO control, they  reveal a fundamental fact in the dynamic PTO control considered in many previous works, e.g.\cite{sergiienko2022comparison,tahir2023latching,
park2016active,jain2022limiting}. The fact is that the leading and lagging reactive PTO forces are indispensable for the transient operation and control of WEC at $\omega \neq \omega_0$. Physically, the reactive PTO force can be easily induced in the generator by the $L$ and $C$ devices as shown above and controlled by controlling $i_C(t)$ and $i_L(t)$. However, this has not drawn much attention in the current literature. 

The presented results are limited to the single-phase 2-pole PMLG, single frequency sinusoidal ocean wave, and steady state operation of WEC. Their extensions to multi-phase multi-pole linear and rotary generators, multi-frequency and broadband waves, and transient operation and control of WEC are currently being investigated. 

\section{Conclusion}
This paper has presented a novel $LC$-tuned WEC and its complete closed loop system model. Three quantitative rules have been derived from the closed loop system model to tune the reactive PTO forces produced by the $LC$ device, for the WEC to resonate at $\omega \neq \omega_0$ and hence generate maximal electric power from ocean waves with variable frequency. Mathematical analysis of the WEC and tuning rules has led to the analytical and quantitative descriptions of the WEC's mechanical power absorption, active and reactive electric power generation and power factor, optimal electric load $R$, and capacity requirements of the PMLG and $LC$ device. 
Simulation results have shown the effectiveness and advantages of the new WEC and verified the analysis results. 
The implications, limitations, and further extensions of the presented results have been discussed.








}}
\bibliographystyle{unsrt}

\end{document}